%% file: nisati-lhcp2014.tex
\documentclass[10pt]{article}
\usepackage{graphicx}
\usepackage{lineno}

\def\Title#1{\begin{center} {\Large #1 } \end{center}}
\def\Author#1{\begin{center}{ \sc #1} \end{center}}
\def\Address#1{\begin{center}{ \it #1} \end{center}}

\newcommand\pubblock{\rightline{\begin{tabular}{l} Proceedings of the Second Annual LHCP\\ \pubnumber\\
         \pubdate  \end{tabular}}}

\newenvironment{Abstract}{\begin{quotation} \begin{center} 
             \large ABSTRACT \end{center}\bigskip 
      \begin{center}\begin{large}}{\end{large}\end{center} \end{quotation}}

\newenvironment{Presented}{\begin{quotation} \begin{center} 
             PRESENTED AT\end{center}\bigskip 
      \begin{center}\begin{large}}{\end{large}\end{center} \end{quotation}}

\input econfmacros.tex

\usepackage{color}
\usepackage{ws-higgs} 

\newcommand\pubnumber{ }

\newcommand\pubdate{\today}

\def\affiliation{
Istituto Nazionale di Fisica Nucleare \\
P.le A. Moro 2, Rome, 00185, Italy }

\begin{document}

\large
\begin{titlepage}
\pubblock

\vfill
\Title{  Experimental Summary  }
\vfill

\Author{ Aleandro Nisati  }
\Address{\affiliation}
\vfill
\begin{Abstract}

High-quality results have been produced with the first Large Hadron Collider
run on high-p$_T$, heavy flavour and heavy
ion physics. These results, as well as the most recent data analyses from
Tevatron, have been presented and discussed at the LHCP2014 conference.
A selection of some of them is summarised in this paper, with care
to those that stimulated interesting discussions during this event.

\end{Abstract}
\vfill

\begin{Presented}
The Second Annual Conference\\
 on Large Hadron Collider Physics \\
Columbia University, New York, U.S.A \\ 
June 2-7, 2014
\end{Presented}
\vfill
\end{titlepage}
\def\thefootnote{\fnsymbol{footnote}}
\setcounter{footnote}{0}
%

\normalsize 


\section{Introduction}

During the conference so many interesting talks and nice results have
been presented, that would be impossible to properly summarize them in
a single paper.
Therefore, a selection is here summarised, with particular 
attention to those that stimulated interesting discussions during 
this workshop.

\section{Electroweak and QCD physics}

A very wide set of studies on Standard Model (SM) have been presented and 
discussed in this conference. These range from QCD jet measurements to 
measurements of photons, leptons, single vector boson and vector boson pair 
production at the Large Hadron Collider (LHC) and Tevatron. 
For single inclusive object studies, the overall  uncertainty on measured 
quantities is dominated by the experimental and theory systematics. 
On the contrary, boson pair measurements 
are almost equally limited by statistical and systematic uncertainties.
In all cases, Monte Carlo predictions show a good agreement with 
next-to-leading order (NLO) or next-to-next-to-leading order (NNLO) calculations.

Vector boson pair production is an important physics process as it represents
one of the most relevant background source to new physics searches.
Furthermore, it allows the study of the gauge boson couplings, probing
the realization in nature of anomalous couplings as predicted by theories
beyond SM. Quartic anomalous couplings (AQGC) have been searched
for at Tevatron and LHC, studying for example the production of 
WW$\gamma$ and WZ$\gamma$ events, as reported in reference 
\cite{Chatrchyan:2014bza}.

\begin{figure}[!h]  
\begin{center}
\includegraphics[width=0.35\linewidth]{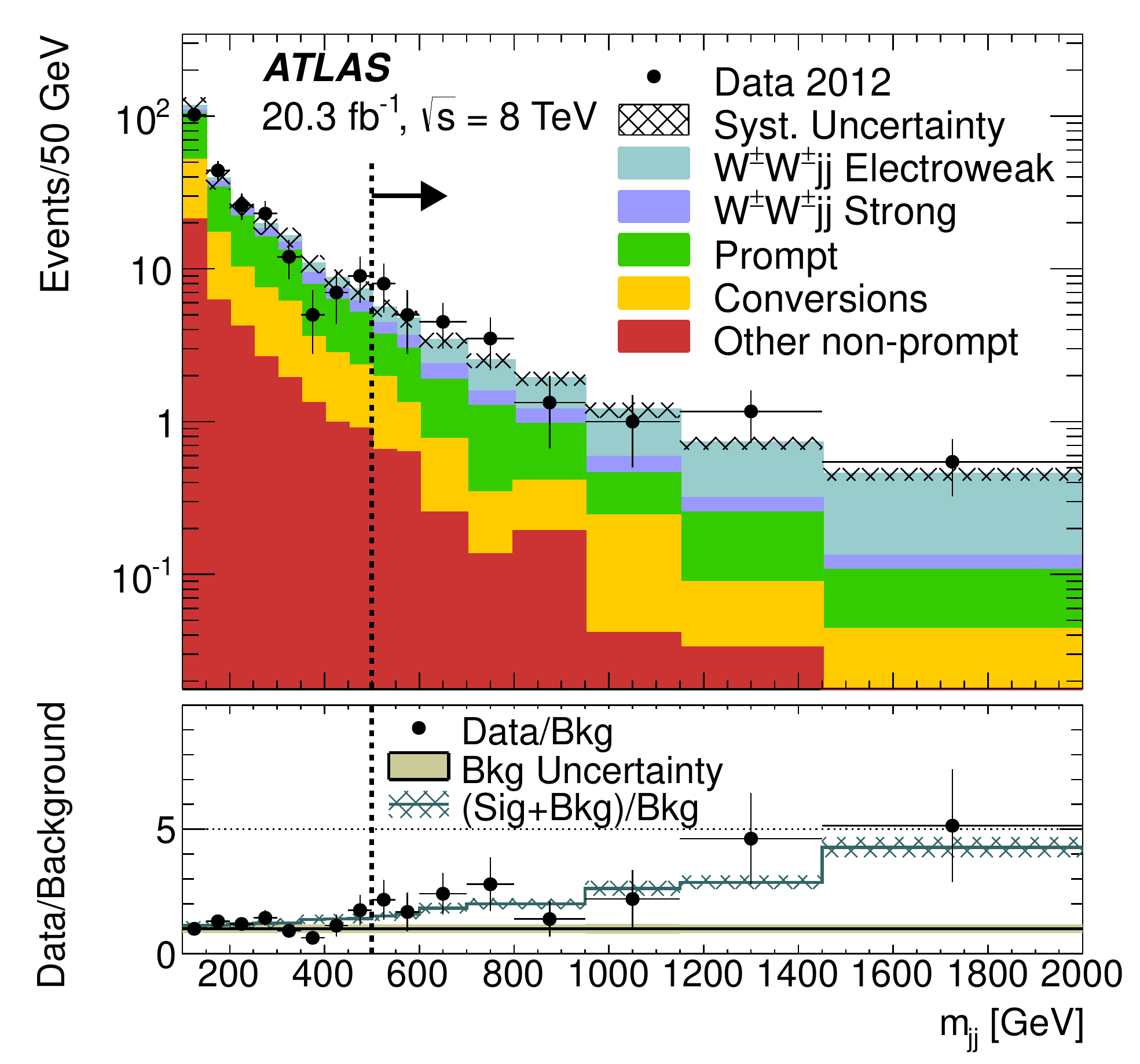} 
\includegraphics[width=0.45\linewidth]{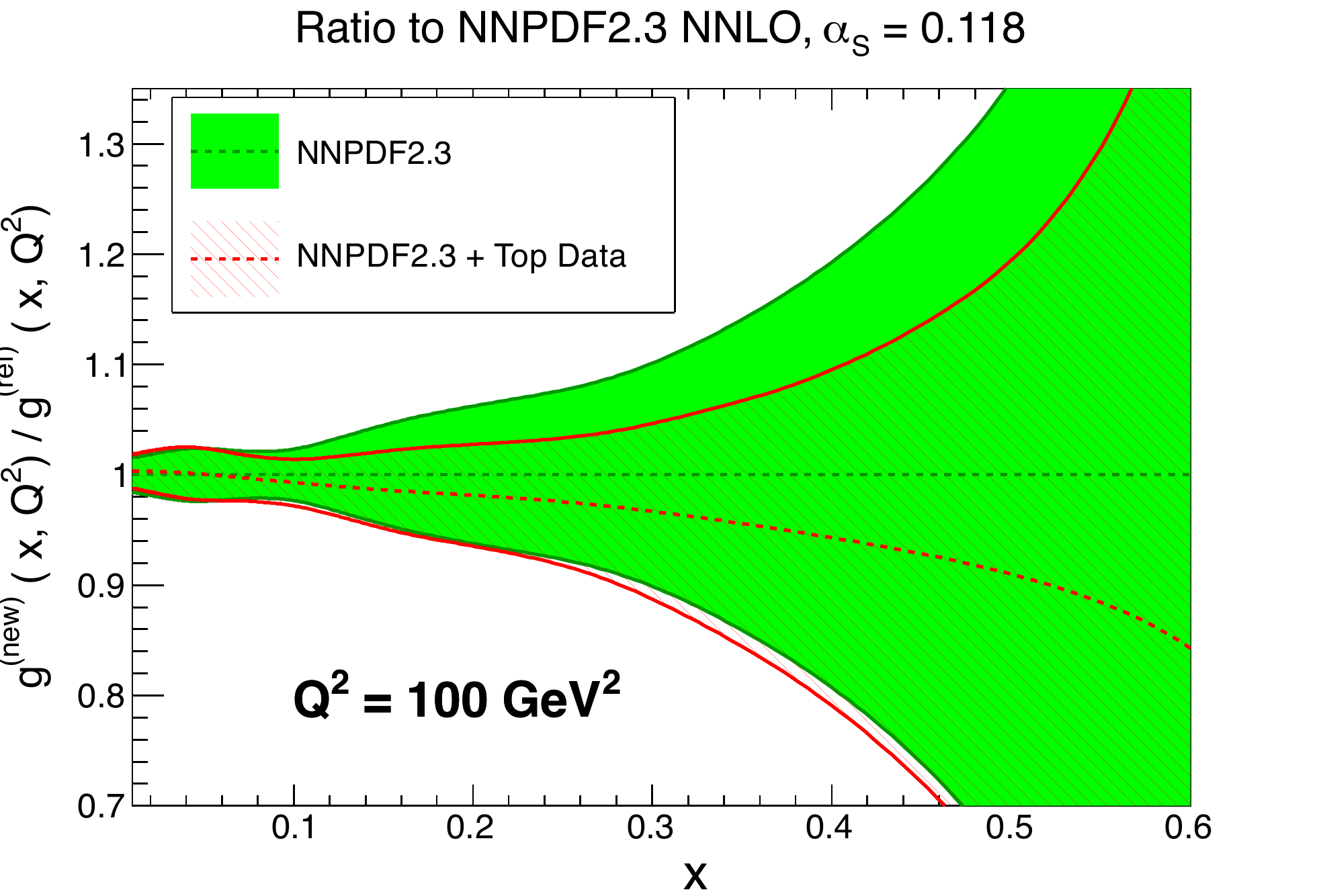} 
\caption{Left: 
The jet-jet invariant mass m$_{jj}$ distribution for events passing the 
inclusive region selections except for the m$_{jj}$ selection indicated by 
the dashed line. 
The $W^\pm W^\pm jj$ events on the left of the dashed line are mainly
from strong production, while those on the right are produced mainly by
electroweak processes.
The black hatched band in the upper plot represents the 
systematic uncertainty on the total prediction. 
On the lower plot the shaded band represents the fractional uncertainty of 
the total background while the solid line and hatched band represents 
the ratio of the total prediction to background only and its uncertainty. 
The $W^\pm W^\pm jj$ prediction is normalized to the SM expectation.
Right:
The ratio of the NNPDF2.3 NNLO gluon PDF at Q$^2$ = 100 GeV$^2$ between 
the default fit and after including the Tevatron and LHC top quark cross 
section data \cite{Czakon:2013tha}.}
\label{sm:plots}
\end{center}
\end{figure}

The Vector Boson Scattering (VBS) is a key process to probe the nature of
the electroweak symmetry breaking mechanism. It is of paramount importance
to understand whether the SM-like Higgs boson recently discovered at LHC
\cite{Aad:2012tfa}\cite{Chatrchyan:2012ufa} is the only process responsible
of the unitarization of this process, or whether, as predicted by many
beyond-Standard-Model scenarios (BSM), other physics reactions contribute
to the VV (V=W,Z) scattering amplitude. A first measurement of same charge 
$W^\pm W^\pm jj$ vector boson production
processes has been made by ATLAS using 20.3 fb$^{-1}$ of data 
at \rts\ = 8 TeV \cite{Aad:2014zda}. 
In the data analysis, two signal regions have been
defined, one loose (inclusive) dedicated to measure the QCD production, 
and the second to extract the electroweak contribution; 
see figure \ref{sm:plots} (left).
An excess of events has been
found in both regions, providing 4.5 and 3.6 standard deviations (s.d.) evidence
for QCD and electroweak production, respectively.

The knowledge of parton distribution functions (PDFs) represents an important 
limitation to the measurement of 125 GeV Higgs boson physics properties, in
particular for the couplings to elementary particles. In fact, they
contribute together with the uncertainties associated to the factorization
and renormalization scale, to an uncertainty that is already comparable to
the experimental uncertainty (see also section \ref{higgs}).
With increasing integrated luminosity at the LHC,
the overall systematic uncertainty will be quickly dominated by theory errors.
This represents an important motivation for the improvement of
the accuracy of PDF measurements in the near future. Though recently proposed 
electron-hadron experiments would be in best position to extract PDF 
information, Tevatron and  LHC data can provide important constraints to PDFs. 
Studies in this direction already started, and an example is available in 
reference \cite{Czakon:2013tha}: figure \ref{sm:plots} (right) shows the 
reduction obtained in an analysis of gluon PDFs where the measured production 
cross sections at Tevatron and LHC atre used.

\section{Top quark physics}

\begin{figure}[!h]  
\begin{center}
\includegraphics[width=0.45\linewidth]{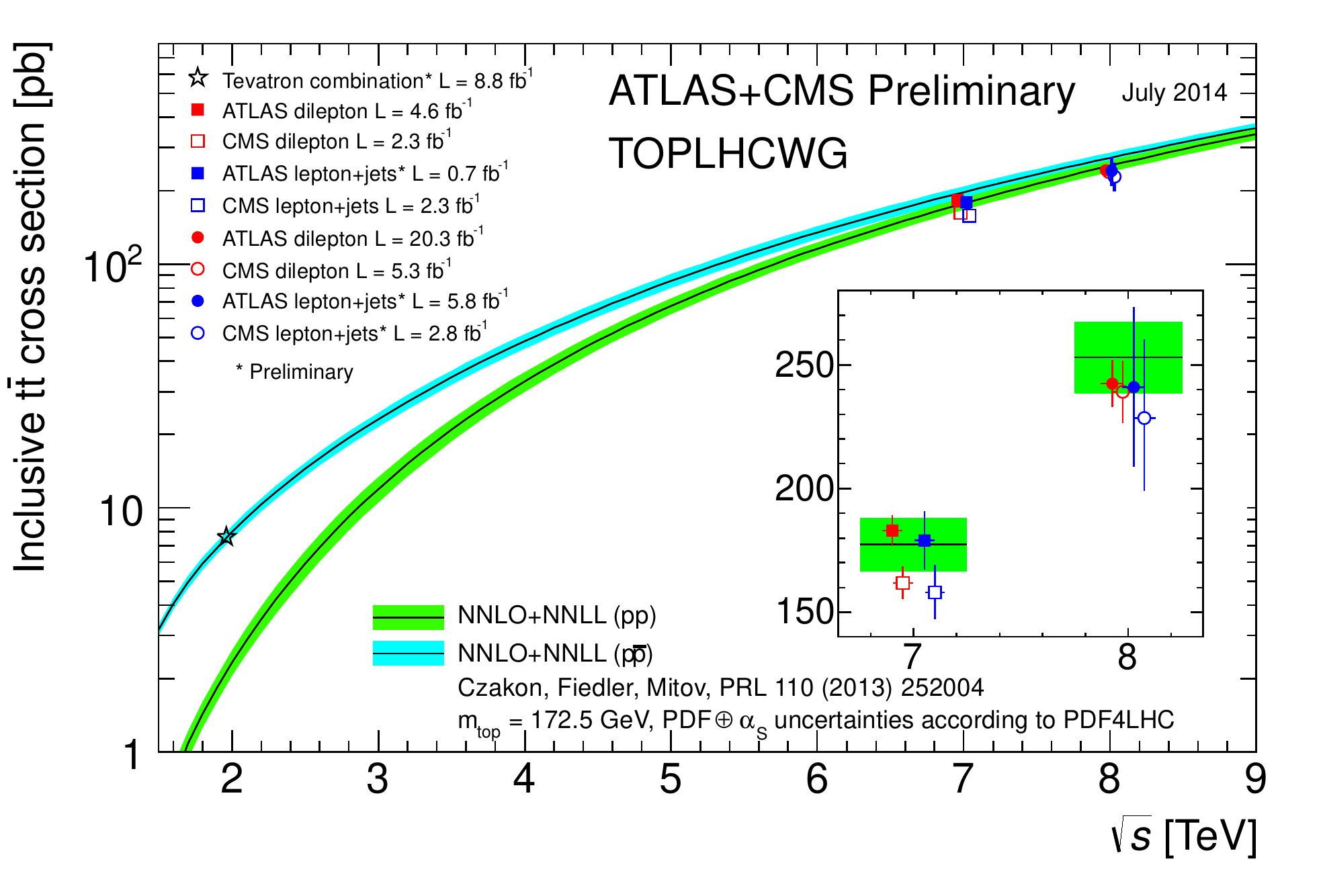} 
\includegraphics[width=0.48\linewidth]{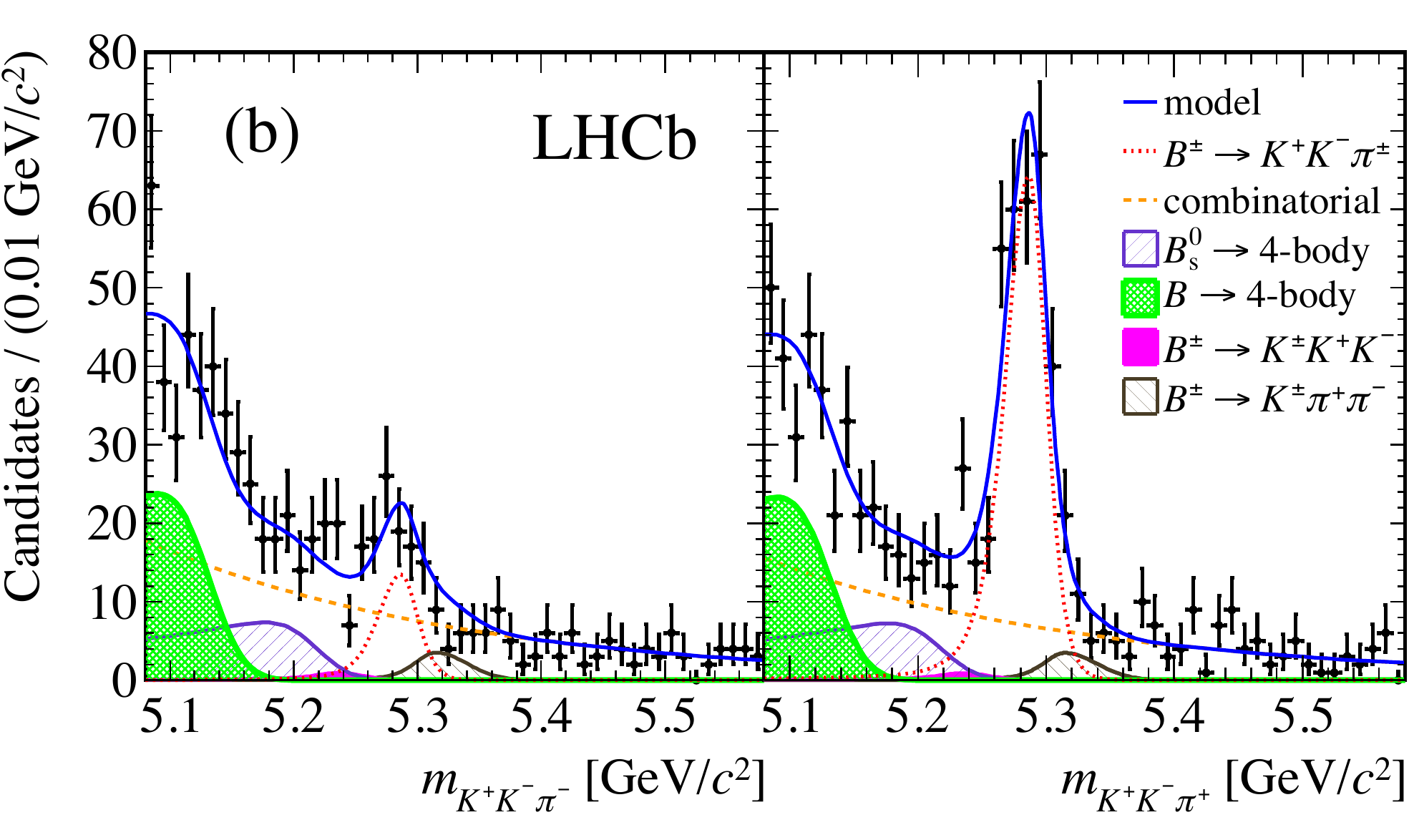} 
\caption{Left: 
Summary of LHC and Tevatron measurements of the top-pair production cross-section as a
function of the centre-of-mass energy compared to the NNLO QCD calculation 
complemented with the next-to-next-to-leading log (NNLL) 
resummation (top++2.0). The theory band represents uncertainties due to
renormalisation and factorisation scale, parton density functions and the strong coupling.
The measurements and the theory calculation is quoted at the top mass $m_{t}$=172.5 GeV
(measurements made at the same centre-of-mass energy are slightly offset for clarity). 
Right: Invariant mass spectra of $B^\pm\rightarrow K^+ K^- \pi^\pm$ 
decays in the region $m^2_{K^+ K^-}<1.5$ GeV$^2$. 
The left-hand panel shows the $B^-$ modes and the right-hand panel shows the 
$B^+$ modes. The results of the unbinned maximum likelihood fits are overlaid
\cite{Aaij:2013bla}.}
\label{hf:plots}
\end{center}
\end{figure}

The production cross section of \ttbar\ pairs and single top quark has been
measured at \rts = 1.96 TeV in \pbarp\ and pp collisions up to
8 TeV in the centre-of-mass energy. The global uncertainty is about 8\% (5\%)
at \rts =7 TeV (8 TeV), per experiment (ATLAS, CMS), and 5\% at 
\rts = 1.96 TeV (CDF, D0). The data are in agreement with theory predictions
within experimental and theoretical accuracy. 
A nice \ttbar\ production cross-section summary is shown in figure 
\ref{hf:plots} (left).

One of the most important set of results shown at the conference concerns the
recent studies of single top production at hadron colliders. There are three main
processes for the production of this quark in single-mode: s-channel, t-channel
and Wt channel. In the s-channel, a quark anti-quark annihilate to produce a
top-anti-beauty quark pair; in the t-channel, an initial state quark interacts with
a gluon, emitting a top-anti-beauty quark pair in association with a quark in the
final state; finally, in the Wt channel a b-quark from the sea interacts with a
initial state gluon producing a top quark in association with a W boson. 
It should be noted that the Wt production is not accessible at Tevatron 
because of the small production cross section in proton-anti-proton collisions at 
\rts\ = 1.96 TeV.
The experiment D0 first observed the t-channel process at Tevatron.
At the LHC, the t-channel has been measured by ATLAS and CMS with an overall accuracy
smaller than 15\% \cite{atlas-conf-2014-007}\cite{Khachatryan:2014iya}.
The Wt process has been observed with a significance of 6.1 s.d. by CMS
\cite{Chatrchyan:2014tua} and with 4.2 s.d. by ATLAS \cite{atlas-conf-2013-010}.  
No significant deviation from Standard Model predictions has been observed in any
of the two processes above summarized.
The single top production in the s-channel has been observed
at Tevatron: the combination of the results from CDF and D0 has produced the 
first observation of this process
with a significance of 6.3 s.d.; also in this case, data are in good
agreement with Standard Model predictions within theoretical and experimental
uncertainties \cite{CDF:2014uma}.
On the contrary, this process has not been observed yet at the LHC, 
and upper limits on its production cross-section have been set
\cite{atlas-conf-2011-118}\cite{cms-pas-top-13-009}.

The measured top mass world combination is $m_t$ = 173$\pm$ 0.76 GeV
\cite{ATLAS:2014wva}. The combination has been performed using the BLUE package,
and efforts are ongoing to harmonize the treatment of the systematic uncertainties.
CDF and D0 reported the final top mass measurements, $m_t$ = 173.16$\pm$0.57(stat)$\pm$0.74(syst) GeV) 
and $m_t$ = 174.98$\pm$0.58(stat)$\pm$0.49(syst) GeV respectively, while CMS has updated previous
measurements ($m_t$ = 172.22$\pm$0.14(stat)$\pm$0.72(syst) GeV). 

One of the most discussed topics during this conference on top physics was certainly
the understanding of the relation between the top mass measured experimentally
in hadron collisions to the pole mass, which enters the Standard Model prediction of
processes involving top quarks.
A nice overview of the problem has been given, and a discussion on how to evaluate the
uncertainty on the connection between the measured quantity and the pole mass used
in the theory \cite{moch:lhcp2014} has been made. In summary, the top quark mass
uncertainty in hadron colliders $\Delta m_t(th)$ is given by the combination of 
a term $\Delta m_t(MC \rightarrow MSR)$ due to the conversion from the MC mass to
the ``short-range" mass (MSR), and a offset $\Delta m_t(MSR \rightarrow pole)$
that converts the short-range mass to the pole mass \cite{Moch:2014tta}. 
In the report given at the conference, 
$\Delta m_t(MC \rightarrow MSR) \sim \pm$0.7 GeV and
$\Delta m_t(MSR \rightarrow pole) \sim +$0.5 GeV.

\section{Heavy flavour physics}

The non-invariance of the combined asymmetry of charge conjugation and parity, 
known as CP violation, is described in the Standard Model by the 
Cabibbo-Kobayashi-Maskawa quark-mixing matrix 
\cite{Cabibbo:1963yz}\cite{Kobayashi:1973fv}.
Charmless decays of B mesons to three hadrons offer the possibility 
to investigate CP asymmetries that are localized in phase space, 
as these decays are dominated by intermediate two-body resonant states.
Moreover, the $B^\pm\rightarrow K^+ K^- \pi^\pm$ decay is interesting because
$s\bar s$ contributions are strongly suppressed. The LHCb experiment has
reported measurements of the inclusive $CP$-violating asymmetries for
$B^\pm\rightarrow \pi^+ \pi^- \pi^\pm$ and $B^\pm\rightarrow K^+ K^- \pi^\pm$
decays. For the first time, an evidence for this asymmetry with a significance 
of 3.2 s.d. has been found \cite{Aaij:2013bla}. Figure \ref{hf:plots} (right)
shows the rate asymmetry of $B^\pm\rightarrow K^+ K^- \pi^\pm$ decays.

This evidence, along with the recent evidence for $CP$ violation in the
$B^\pm\rightarrow K^\pm \pi^+ \pi^-$ and $B^\pm\rightarrow K^\pm K^+ K^-$
decays \cite{Aaij:2013sfa}
and recent theoretical developments, indicates new mechanisms for $CP$
asymmetries, which should be included in models for future amplitude
analyses of charmless three-body decays.

Violations of $CP$ symmetry are predicted to be small in charm decays, 
but could be enhanced in the presence of non-SM physics. 
Direct CP violation arises when two or more amplitudes with different
weak and strong phases contribute to the same final state. 
This is possible in singly Cabibbo-suppressed $D^0\rightarrow K^- K^+$
and $D^0\rightarrow \pi^- \pi^+$ decays where significant 
penguin contributions can be expected \cite{Bobrowski:2010xg}.
To date, $CP$ violation in charm decays has not been established experimentally.
$CP$ asymmetries have been recently investigate by the LHCb collaboration
measuring the asymmetry $\Delta A_{CP}$, for the two decays above.
No observation of $CP$-violation has been shown at the level of
$\Delta A_{CP} \sim 10^{-3}$ \cite{Aaij:2014gsa}.
This represents the most precise measurement of time-integrated $CP$
asymmetries in the charm sector from a single experiment.

Lots of studies have been discussed on heavy-flavour spectroscopy.
The observation of the particle $Z(4430)$ by LHCb has been shown with
a signicance in excess of 14 s.d.; this object was already observed
by the experiment Belle in 2008, but with no conclusion from
the experiment BaBar. Also, evidence for the decay $X(3872)
\rightarrow \psi(2S)\gamma$ has been presented. $B_c$ physics studies
have produced many results, in particular, LHCb has performed 
the most precise measurement of the $B^\pm_c$ hadron lifetime, 
using the decay channel 
$B^\pm \rightarrow J/\psi ~\mu^\pm \nu$: 
$\tau = (509\pm 8 (stat) \pm 12 (syst)$) fs.

\section{The 125 GeV Higgs boson and BSM Higgs boson searches}
\label{higgs}

One of the most discussed subjects during this conference was the final ``125 GeV"
Higgs boson \cite{Aad:2012tfa}\cite{Chatrchyan:2012ufa}
mass measurement by ATLAS, based on the data collected in the first run 
of LHC.
The measurement is based on the analysis of inclusive channels
\Hgg\ and \HZZllll\ and it is has been performed after the new electron and
photon energy calibration using the data from the first LHC run \cite{Aad:2014nim}.
This new calibration relies on improved corrections of the calorimeter
non-uniformities and layer inter-calibration based on electron/photon/muon data,
individual electromagnetic cluster energy correction using multi-variate methods, 
and an improved calorimeter energy-scale and energy-resolution determination using
high statistics of \Zee\ decays.

\begin{table}[t]
\begin{center}
\begin{tabular}{l|ccc}  
channel     &  ATLAS-old & ATLAS-Run1 &  CMS           \\ \hline
 \Hgg\      & 126.8 $\pm$ 0.2 $\pm$ 0.7
            & \textbf{125.98 $\pm$ 0.42 $\pm$ 0.28}
            & 125.4 $\pm$ 0.5 $\pm$ 0.6    \\
 \HZZllll\  
            & 124.3 $^{+0.6}_{-0.5}$ $^{+0.5}_{-0.3}$
            & \textbf{124.51 $\pm$ 0.52 $\pm$ 0.04}
            & 125.6 $\pm$ 0.4 $\pm$ 0.2   \\ \hline
 Combination & 125.5 $\pm$ 0.2 $^{+0.5}_{-0.6}$
             & \textbf{125.36 $\pm$ 0.37 $\pm$ 0.18}
             & 125.7 $\pm$ 0.3 $\pm$ 0.3 $^{(*)}$ \\  \hline
\end{tabular}
\caption{Higgs boson mass measurements, in GeV, from ATLAS and CMS
at the time of this conference. As long as ATLAS is concerned, final results from
Run1 analysis are reported, compared to those previously available.
For each measurement the first error represents the statistical uncertainty,
while the second is the overall systematic uncertainty.
$^{(*)}$ The CMS combined mass has not been updated with the new measurement from
the \HZZllll\ channel, and the value available in reference \cite{cms-pas-hig-13-005}
has been reported.}
\label{tab:higgs-mass}
\end{center}
\end{table}

The results, together with a new categorization 
of the  \Hgg\ final states, based on the full 2011 and 2012 data sets, are summarized
in Table \ref{tab:higgs-mass}\cite{Aad:2014aba}, together with the most up do date
results from CMS \cite{cms-pas-hig-13-001}\cite{Chatrchyan:2013mxa}\cite{cms-pas-hig-13-005}. 
At the time of the writing of these proceedings, CMS has reported
a new Higgs boson mass measurement where the latest results on \Hgg\
and \HZZllll\ are used in the combination \cite{cms-pas-hig-14-009}. 
The consistency between the ATLAS mass measurements in the diphoton channel and in the
4-lepton channel is about 2 standard deviations. Furthermore, 
the systematic uncertainty in the ATLAS combined measurement has decreased 
by a factor 3 thanks to the more accurate electron and photon calibration.

\begin{figure}[!h]  
\begin{center}
\includegraphics[width=0.38\linewidth]{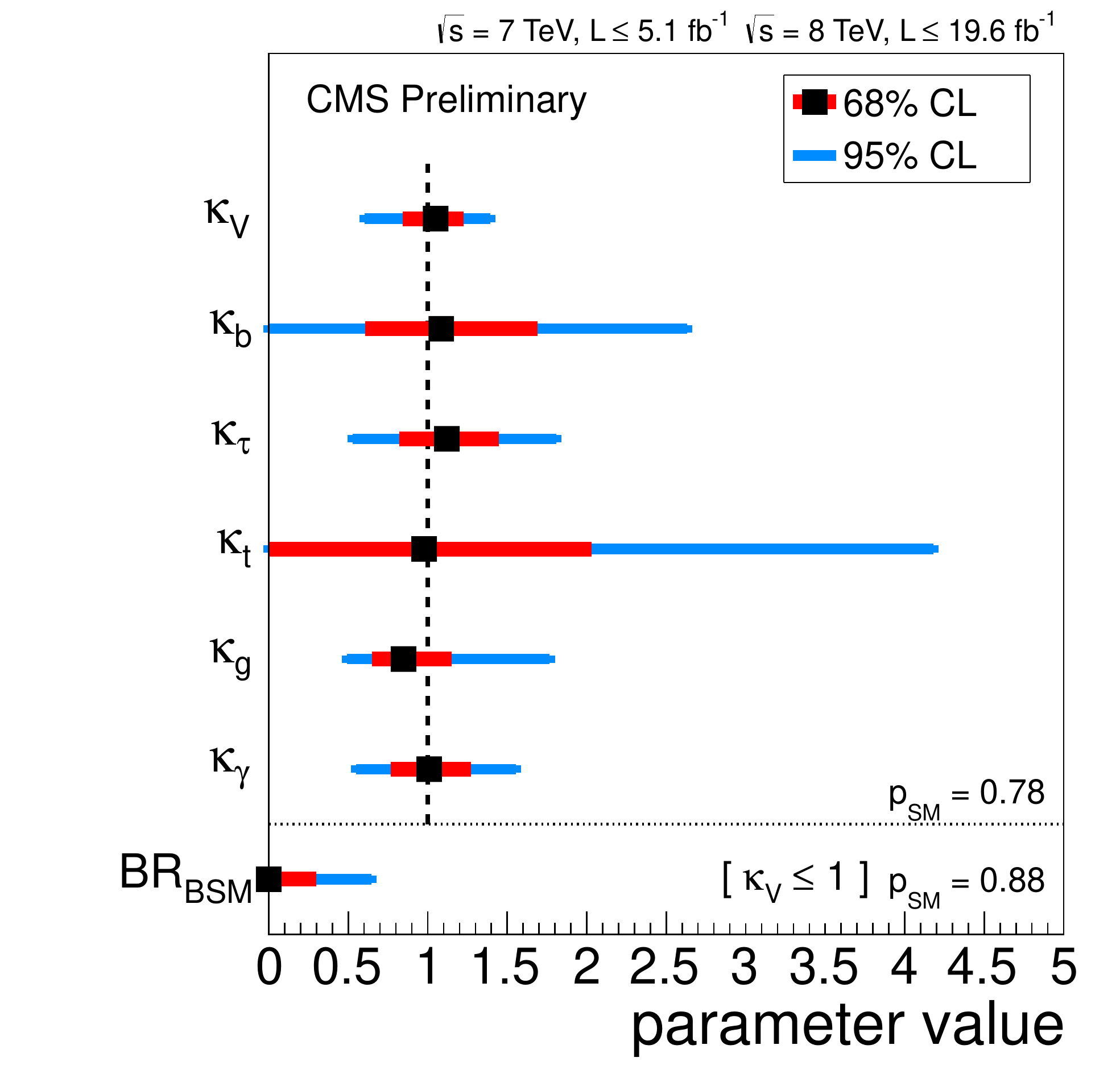} 
\includegraphics[width=0.45\linewidth]{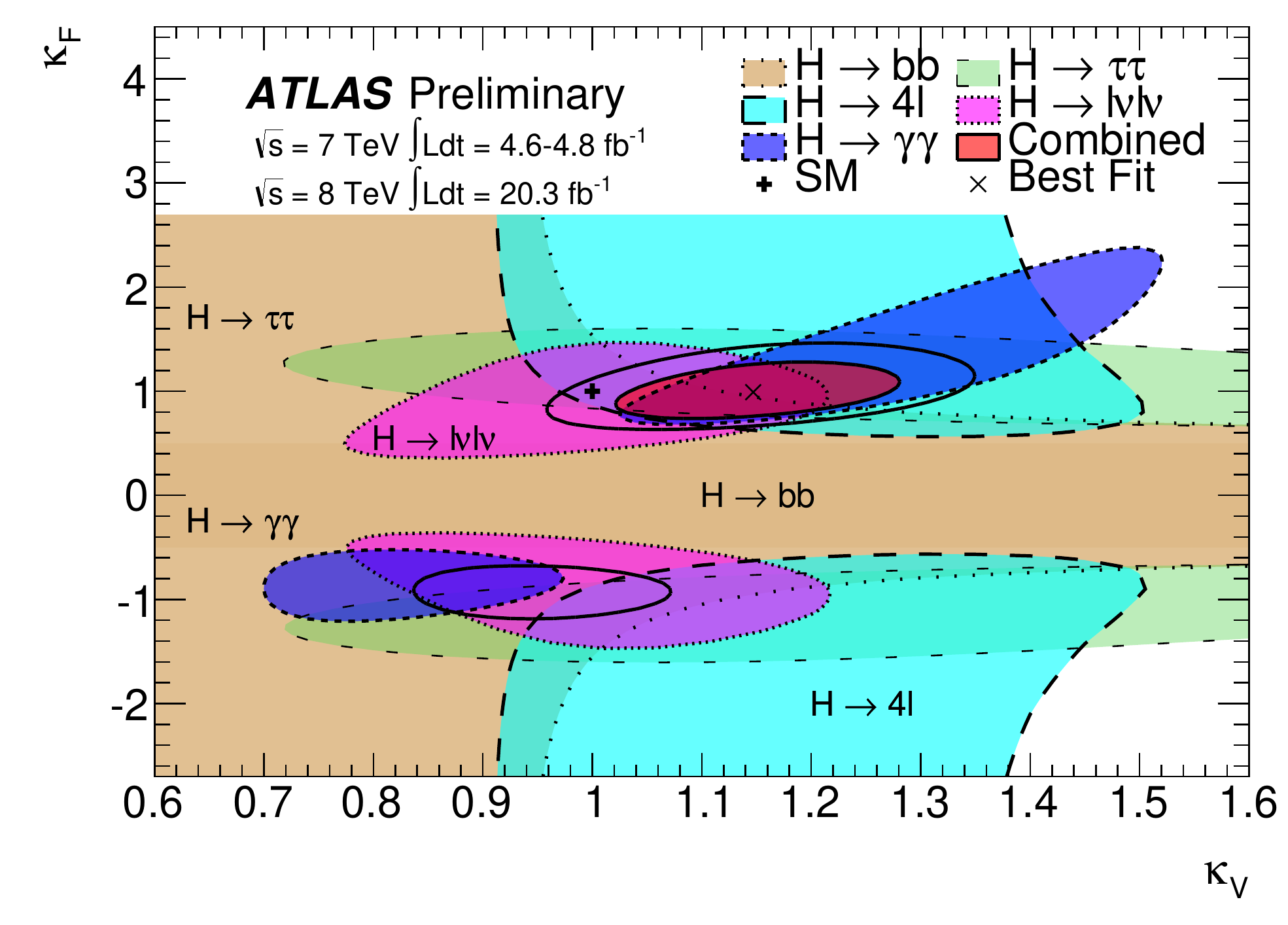} 
\caption{Left: 
Summary of the fits for deviations in the coupling for the benchmark models in reference 
\cite{Heinemeyer:2013tqa}. For each model, the best fit values of the most interesting parameters are shown, with the corresponding 68\% and 95\% CL intervals. The list of parameters for each model and the numerical values of the intervals are provided 
in reference \cite{cms-pas-hig-13-005}.
Right: Results of fits for the 2-parameter benchmark model defined in Section 5.2.1 
of reference \cite{atlas-conf-2014-009} that probe different coupling strength scale
factors for fermions and vector bosons,$\kappa_V$ and $\kappa_F$, 
assuming only SM contributions to the total 
width, overlaying the 68\% C.L. contours derived from the individual channels and their 
combination.}
\label{higgs:plots}
\end{center}
\end{figure}

Impressive results have been produced also from the study of the Higgs boson
couplings to elementary particles. This includes also the search for this scalar 
in two-fermion final states, in particular \Hbb\ and \Htau\ decays. Evidence 
of the production and decay of this particle in these final states has been
shown by ATLAS and CMS with a  significance of 3.7 s.d. \cite{atlas-conf-2014-009}
and 4.4 s.d. \cite{Chatrchyan:2014vua}, respectively.
The signal strength $\mu$, which is the ratio between the measured Higgs 
production and decay rate to the one expected from the Standard Model, 
has been measured by ATLAS to be 
$\mu = 1.30 \pm 0.12$(stat) $^{+0.14}_{-0.11}$(syst); 
theory uncertainty (mainly from parton distribution functions, PDFs, and QCD 
factorization and renormalization scale) are comparable to those from 
experimental sources. Similar results have been shown
by CMS. The signal rate from different production mechanisms and different decay channels
have been used to extract the Higgs boson couplings to elementary 
fermions and bosons.
Several coupling models have been studied, following the prescription and the
recommendations published in \cite{Heinemeyer:2013tqa}. The results have been 
obtained assuming that only Standard Model particles contribute to the total 
natural width, and to the production and decay loops. 
These findings have shown good agreement with Standard Model predictions. 
An example of a particular model is shown in figure \ref{higgs:plots} (left),
where 6 coupling parameters are studied simultaneously.
In this particular model it is possible to see that couplings are measured with an
accuracy of the order 20\% - 50\%.
The analysis of events with a Z-boson produced in association with large missing
transverse energy can be used to search for ZH pairs, where the Higgs boson decays
to ``invisible" final states. Limits have been set assuming that the coupling of the
125 GeV scalar with the W and Z bosons does not exceed the SM predictions, and are
shown in the last row of the figure. 

In a different model, only two parameters are left free in the fit: $\kappa_V$ 
($\kappa_V = \kappa_W = \kappa_Z$) and
$\kappa_F$ ($\kappa_F = \kappa_t = \kappa_b = \kappa_\tau = \kappa_g$), 
that measure the Higgs boson couplings to elementary bosons and fermions,
respectively, in units of the Standard Model prediction. The result of this fit is
shown in figure \ref{higgs:plots} (right), for the individual channels and for their
overall combination.

Studies of quantum numbers of the 125 GeV Higgs boson have been presented and
discussed. In particular, the spin-parity hypothesis $J^P = 0^+$ has
been tested against alternative hypotheses such as $J^P = 0^-, 1^+, 1^-$, as well
as against some graviton-inspired $J^P = 2^+$ and $2^-$ scenarios \cite{Bolognesi:2012mm}.
Results have been shown by ATLAS, CMS and D0. All tested alternative spin hypotheses
appear disfavored compared to the hypothesis $J^P = 0^+$ at more than 97.8\% confidence
level.

A direct measurement of the Higgs boson natural width $\Gamma_H$, predicted by the
Standard Model to be $\Gamma_H = 4.15$ MeV for m$_H$ = 125.6 GeV,  is not possible
at LHC, because of the experimental mass resolution. However, it has been shown 
\cite{Kauer:2012hd}\cite{Caola:2013yja} that it is possible to constraint 
the size of $\Gamma_H$ measuring the ZZ production rate for invariant mass values
away from the resonance. Using 4-lepton final states and 2-lepton + missing transverse
energy final states, CMS presented an upper limit $\Gamma_H \leq$ 22 MeV at 95\%
confidence level \cite{Khachatryan:2014iha}; 
a similar result has been presented by ATLAS \cite{atlas-conf-2014-042}.

Beyond Standard Model physics in the Higgs sector has been largely probed following
two orthogonal paths: performing direct searches for partners of recently 
discovered scalar, predicted by a large number of models that extend 
the Standard Model, and testing these
models using the available data on Higgs boson rates.
No evidence of new objects associated to the newly discovered particle has been found
in searches performed at LHC and Tevatron 
\cite{cms-pas-hig-13-021}\cite{Aad:2012cfr}\cite{Aaij:2013nba}\cite{Abazov:2011up}\cite{Aaltonen:2012zh}\cite{Chatrchyan:2013qga}.

\section{Searches beyond Standard Model}

\begin{figure}[!h]  
\begin{center}
\includegraphics[width=0.42\linewidth]{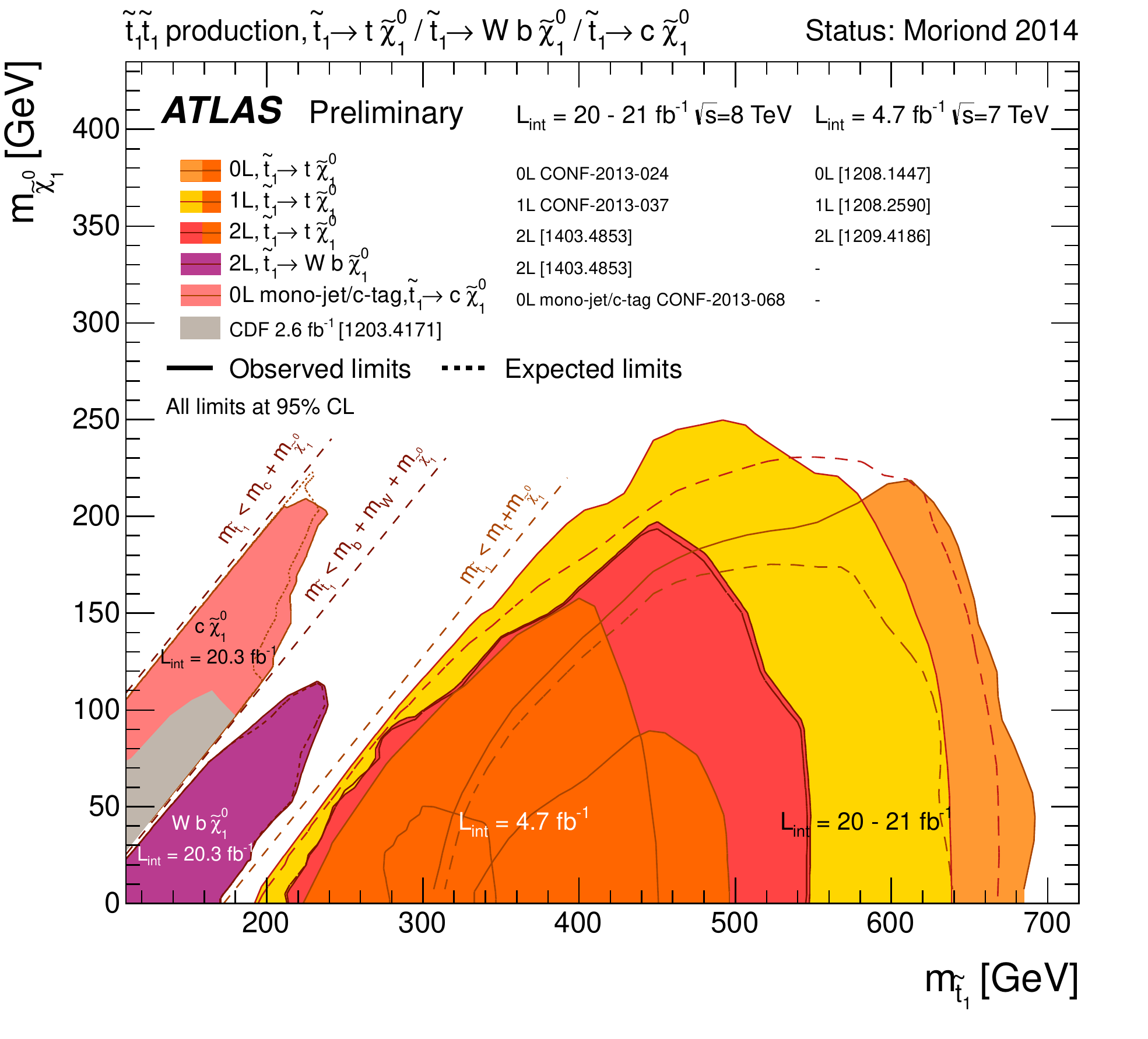} 
\includegraphics[width=0.38\linewidth]{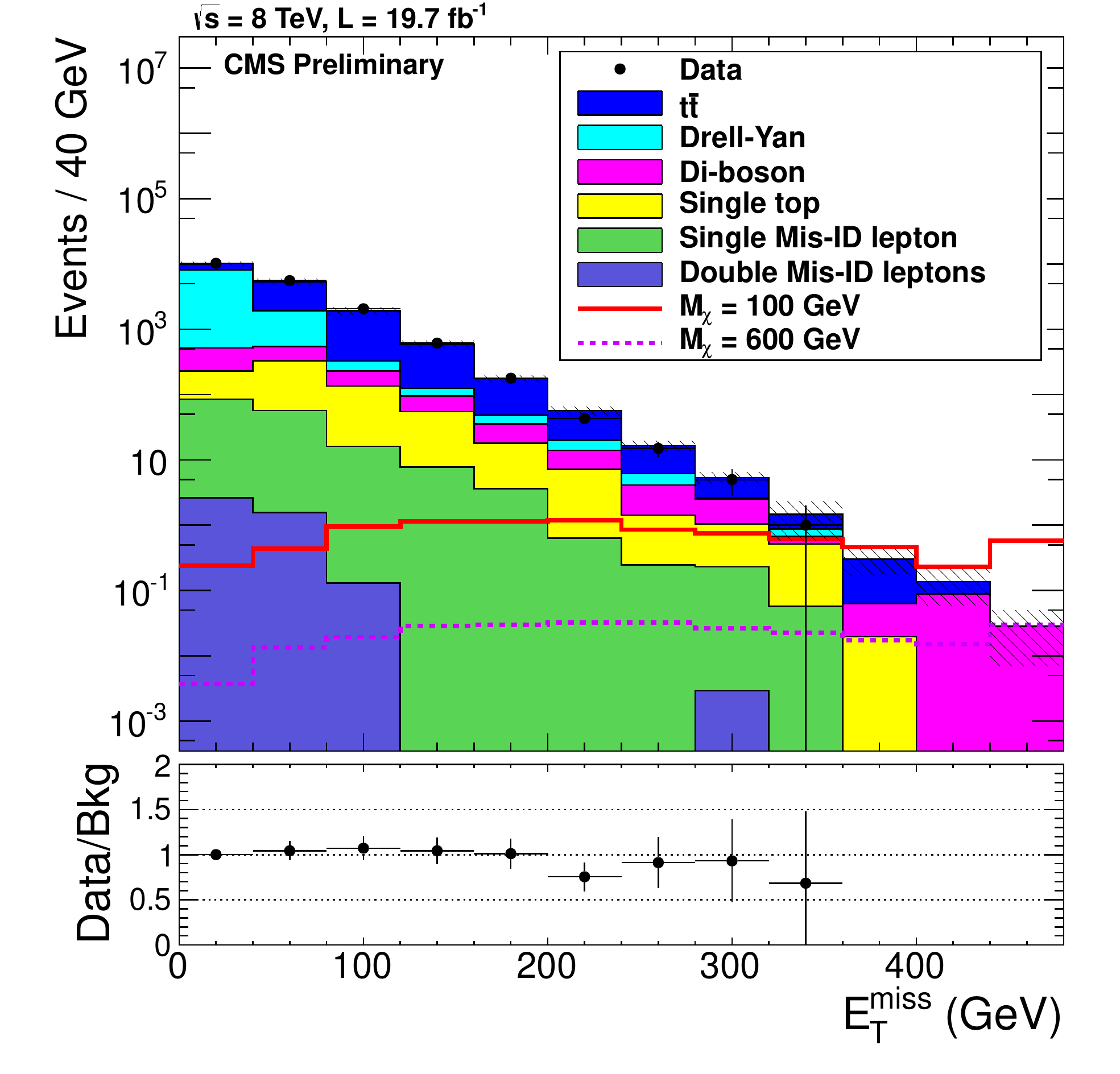} 
\caption{Left:
summary of the dedicated ATLAS searches for top squark pair
production based on 20-21 fb$^{-1}$ of pp collision data taken at
\rts\ = 8 TeV, and 4.7 fb$^{-1}$ at \rts\ = 7 TeV. For more details,
see \cite{atlas-susy}.
Right:
Distributions of the E$_T^{miss}$ after applying selection cuts. 
In this figure, two simulated dark matter signals with mass M$_X$ = 100 GeV
and 500 GeV and the interaction scale M$_*$=100 GeV are included for comparison.
The shaded region represents the total uncertainty of the background 
prediction.  
The error on the data-over-background ratio takes into account both the 
statistical uncertainty of data and the total uncertainty of background
\cite{cms-pas-b2g-13-004}. The last bin includes the overflow. }
\label{bsm:plots}
\end{center}
\end{figure}

Supersymmetry (SUSY) provides an elegant solution to cancel the quadratic 
mass divergences that would accompany a Standard Model Higgs boson by
introducing supersymmetric partners of all SM particles, such as a scalar
partner of the top quark, the top squark $\tilde{t}$. 
The viability of the of SUSY as
a scenario to stabilize the Higgs potential and to be consistent with 
electroweak naturalness is tested by the search for $\tilde{t}$ below the
TeV scale.
An example of a nice summary of direct top squark at LHC with the ATLAS
detector is available in figure \ref{bsm:plots} (right).
Top squarks with masses between ~ 200 GeV and ~ 700 GeV decaying 
to an on-shell t-quark and a neutralino are excluded at 95\% 
confidence level for a light neutralino. Similar results have been shown
by CMS, and this is now challenging the naturalness of this theory, that
represents one of its strongest points.

Other extensive searches for new physics objects predicted by SUSY 
implementations have been presented and discussed. 
Particularly important are also the new results
dedicated to the searches for electroweak production of charginos, neutralinos
and sleptons. 
A variety of signatures with leptons, W, Z and Higgs bosons have been
investigated by CMS \cite{Khachatryan:2014qwa}, 
while ATLAS has looked to final states with two
leptons and large missing transverse energy \cite{Aad:2014vma}.
No significant excess beyond Standard Model expectations has been observed,
and limits have been set on the masses of charginos and neutralinos in
simplified models.

Many searches on non-SUSY beyond-Standard-Model signatures have been
presented and discussed. These include investigation of final states
of heavy bosons W' and Z' decaying to lepton and heavy quark final states, 
as well as SM boson pairs. Other phenomena such as events produced by
new contact interactions have been probed. No excess of events with respect
to the expected yield has been found, and exclusion limits have been set
on the mass of new particles predicted by several BSM models, or on the
energy scale at which new forces could appear.

Dark matter searches are one of the central subjects of BSM studies
at LHC. Many theories predicts dark matter particles light enough to
be produced by this collider. If produced, these objects would escape
the detector producing a large missing transverse energy. In some particular
implementation, dark matter particle pairs are produced in association
with top-antitop pairs, whose leptonic decays can be used to tag the
events. Figure \ref{bsm:plots} (right) shows the distribution of E$_T^{miss}$
after selections cuts for data and predictions, showing no evidence for
production of new physics in this type of investigations
\cite{cms-pas-b2g-13-004}.

\section{Heavy Ion physics}

Many new - and often unexpected - results have been presented at this
Conference. Quite a few of them are related to p-Pb collision studies,
made with data collected a couple of years ago at $\sqrt{s_{NN}}$ = 2.76 TeV.
Among these, the analysis of the {\em ridge} performed using pp, PbPb and pPb
data, and comparing the respective findings, has shown surprising results.
As long-range correlations are seen in both PbPb and pp collisions, it was
natural to expect a possible effect also in pPb collisions. However, within
a week of data-taking it appeared already clear that the correlations in pPb 
are surprisingly much stronger than in pp collisions \cite{WeiLi:lhcp2014}.
A large set of measurements strongly suggest that the ridge(s) results seem to
indicate that the same collective physics that we think we understand in PbPb is
present also in pp and pP on soft scales.

\begin{figure}[!h]  
\begin{center}
\includegraphics[width=0.35\linewidth]{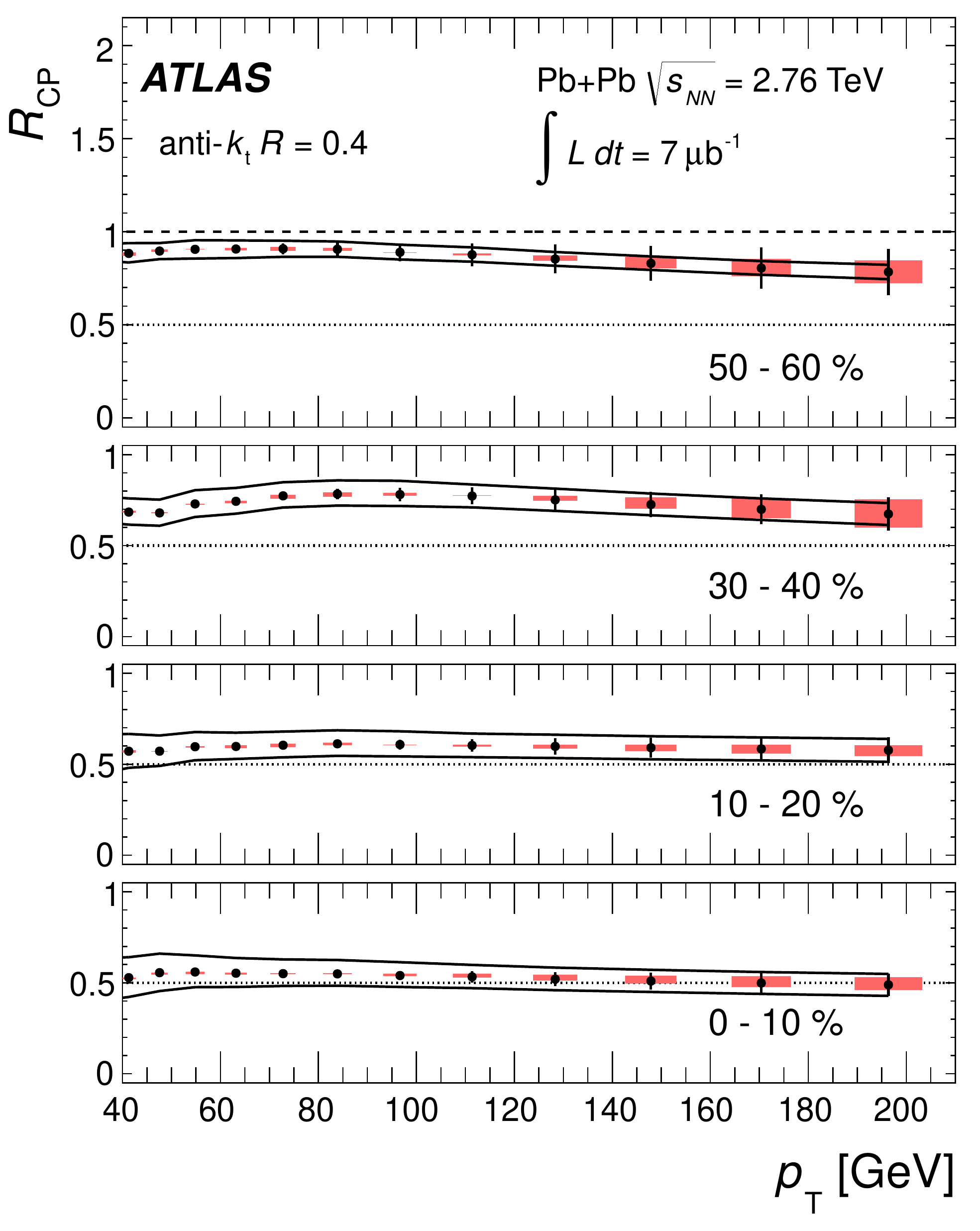} 
\includegraphics[width=0.50\linewidth]{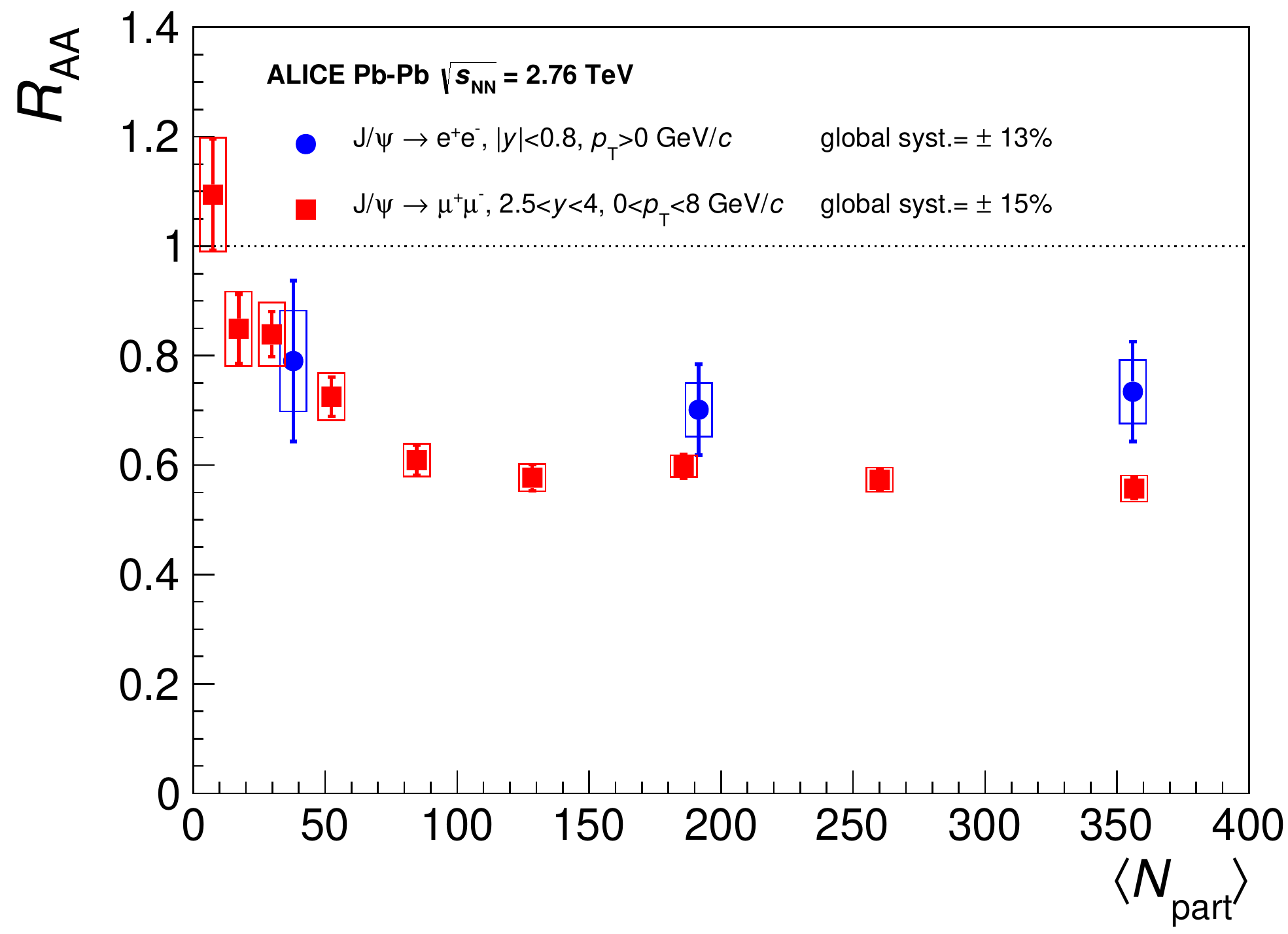} 
\caption{Left: 
Unfolded $R_{CP}$ values as a function of jet \pT\ for 
R = 0.4 anti-kt jets in four bins of collision centrality. 
The error bars indicate statistical errors from the unfolding, 
the shaded boxes indicate unfolding regularization systematic errors 
that are partially correlated between points. 
The solid lines indicate systematic errors that are fully correlated 
between all points. 
The horizontal width of the systematic error band is chosen for 
presentation purposes only. Dotted lines indicate $R_{CP}$ =0.5, 
and the dashed lines on the top panels indicate $R_{CP}$ = 1 \cite{Aad:2012vca}. 
Right: Centrality dependence of the nuclear modification 
factor, $R_{AA}$, of inclusive J$\psi$ production in Pb-Pb collisions
at $\sqrt{s_{NN}}$ = 2.76 TeV, measured at mid-rapidity and at forward
rapidity \cite{Aad:2012vca}.}

\label{hi:plots}
\end{center}
\end{figure}

Another major topic discussed was the measurement of the single jet suppression,
to to the propagation of partons in the hot dense medium of quark-gluon
plasma (QGP) produced in heavy ion collisions.
This is done measuring the jet yield in a given \pT\ and centrality bin, and
rescaling taking into account the number of nucleon-nucleon collisions
expected for that bin. The data analysed are from a sample of inclusive
jets produced in PbPb collisions at $\sqrt{s_{NN}}$ = 2.76 TeV. The ratio
$R_{CP}$ of yields measured at different centralities to the one observed
in the centrality bin 60\%-80\% is determined and plotted as a function 
of the reconstructed jet \pT. A factor $\sim$ 2 suppression in jet rate 
is observed \cite{Aad:2012vca}; see also figure \ref{hi:plots} (left).

Heavy quarks are an important probe of the QGP since they are expected to be 
produced only during the initial stage of the collision in hard partonic 
interactions, thus experiencing the entire evolution of the system. 
It was predicted that in a hot and dense deconfined medium like the QGP,
bound states of charm and anti-charm quarks, i.e. charmonia, 
are suppressed due to the screening effects induced by the high density 
of color charges. The relative production probabilities of charmonium
states with different binding energies may provide important information 
on the properties of this medium and, in particular, on its temperature. 
Among the charmonium states, the strongly bound J$\psi$ is of particular 
interest \cite{Abelev:2013ila}. The behaviour of the nuclear modification
factor,~$R_{AA}$, as a function  of the centrality $<N_{part}>$ shown
in figure \ref{hi:plots} (right) indicates that 
for increasing values of this variable, a constant value of $R_{AA}$ is 
observed, suggesting some type of regeneration processes. 
The charmonium loses the status of ``thermometer"
of the QGP medium, and may play an important role in the the understanding 
of the phase boundary.

\section{Future Prospects}

The LHC machine is in shutdown since March 2013, mainly for 
magnet interconnect repairs, to allow nominal current in the
dipole and lattice quadrupole circuits of the LHC \cite{lhc-1}.
This should bring the collision energy to a value close to \rts\ = 14 TeV.
The instantaneous luminosity will increase up to the nominal value 
$\mathcal{L} = 10^{34}$ cm$^{-2}$s$^{-1}$, and will be doubled with
another machine upgrade planned for 2018.
The LHC will accumulate data corresponding to an integrated 
luminosity of about 300 fb$^{-1}$ by 2022, or so.

To extend its discovery potential, the LHC will need a major upgrade  
to increase its luminosity by a factor of 5 (or more) 
beyond its design value. As a highly complex and optimized machine, 
such an upgrade of the LHC must be carefully studied and requires about 
10 years to implement.

The novel machine configuration, the High-Luminosity LHC (HL-LHC), 
will rely on a number of key innovative technologies, representing exceptional 
technological challenges, such as a few cutting-edge 13 Tesla superconducting magnets,
very compact and ultra-precise superconducting cavities for beam rotation, 
and 300-metre-long high-power superconducting links with zero energy 
dissipation \cite{hl-lhc}\cite{hl-lhc-1}\cite{Rossi}.
 
\begin{figure}[!h]  
\begin{center}
\includegraphics[width=0.44\linewidth]{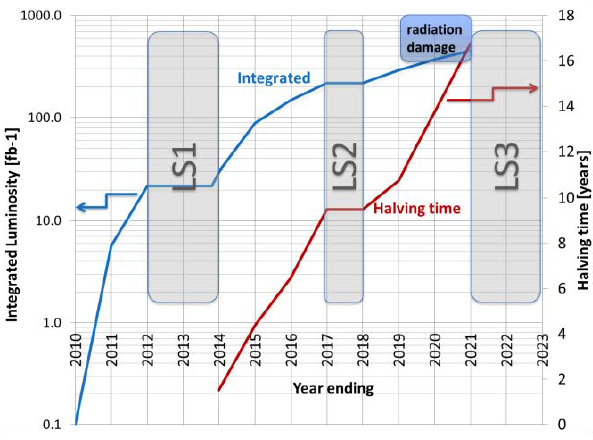} 
\includegraphics[width=0.46\linewidth]{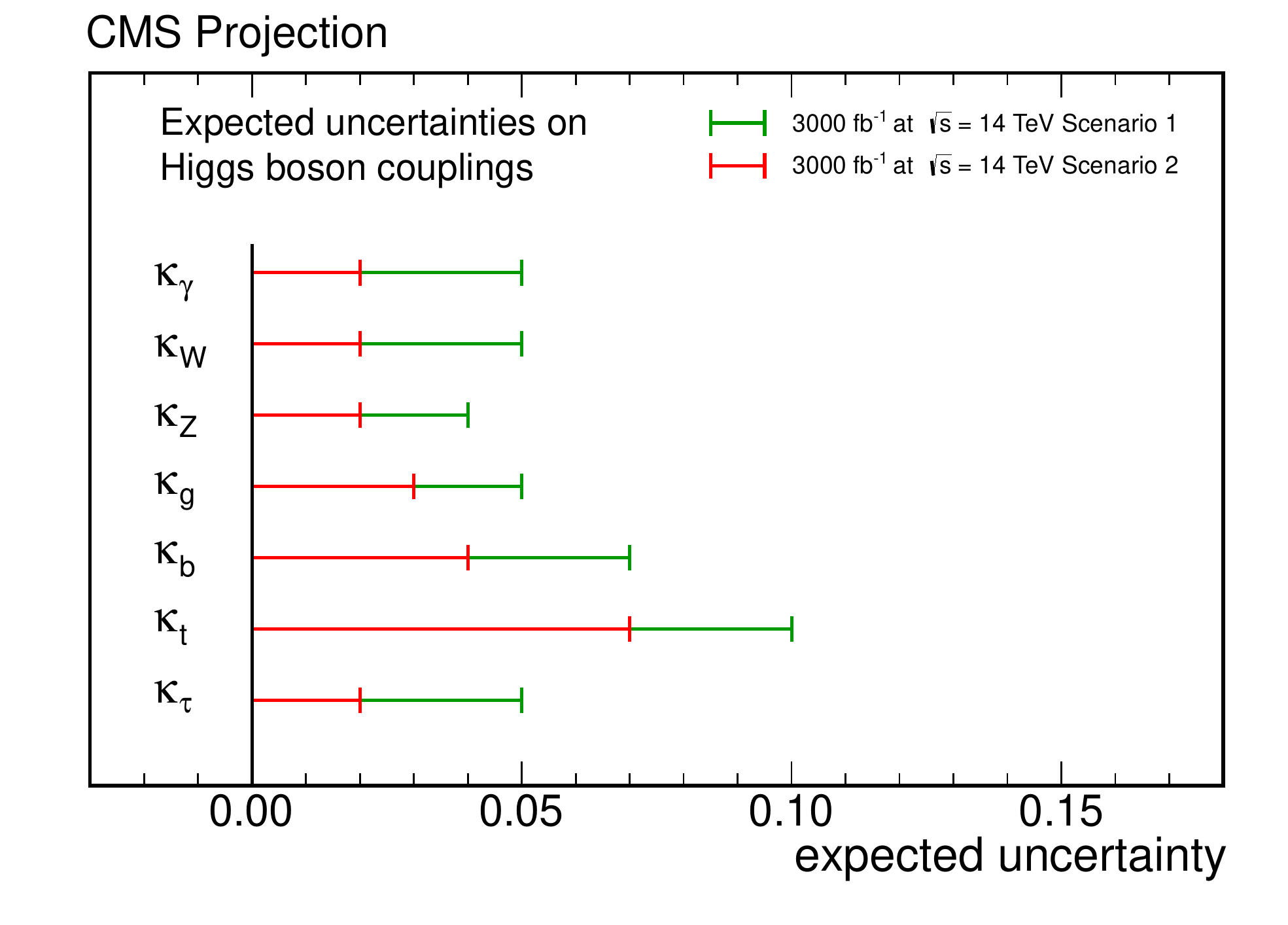} 
\caption{Left: 
Integrated luminosity and running time to reduce by a factor two the statistical
error based on flat luminosity (halving time). Superimposed the three long 
shutdowns and the area where it is expected that radiation damage will 
call for changing of the low-$\beta$ quadrupoles (also called inner triplet).
Right:
 Estimated precision on the measurements of $\kappa_\gamma$, $\kappa_W$,
 $\kappa_Z$, $\kappa_g$, $\kappa_b$ and $\kappa_\tau$. The projections
 assume \rts\ = 14 TeV, a dataset corresponding to 3000 fb$^{-1}$ and 
 that no particles other than those from SM
 contribute to the Higgs boson natural width or to the loops. Furthermore,
 Scenario 1 assumes current experimental and theoretical uncertainties,
 while Scenario 2 assumes that these uncertainties scale with 1/$\sqrt{(L)}$
 (where $L$ is the integrated luminosity) and 1/2, respectively.}
\label{prosp:plots}
\end{center}
\end{figure}

With this final upgrade, the LHC machine should deliver an integrated luminosity
of about 3000 fb$^{-1}$ per experiment (ATLAS and CMS) by $\sim$ 2030,
see figure \ref{prosp:plots} (left).

With this ultimate data set, the HL-LHC will be the unique worldwide facility
to look for rare processes and it will give an unprecedented sensitivity for
a large range of the newly discovered Higgs boson property measurements, as
well as for searches for new particles and precision studies for a wide
set of fundamental physics processes \cite{ECFA-Report}.

Figure \ref{prosp:plots} (right) shows the expected accuracy on the 125
GeV Higgs boson couplings computed by CMS \cite{CMS:2013xfa}; 
ATLAS has shown very similar results \cite{atlas-phys-pub-2013-004}.

\section{Conclusions}

The first run of LHC has shown outstanding performance of the machine and of the 
ALICE, ATLAS, CMS and LHCb detectors. 
High-quality results have been produced in high-p$_T$, heavy flavour and heavy
ion physics. A new boson, compatible within experimental and theoretical
uncertainties with the Standard Model Higgs boson, has been observed
at a mass of about 125 GeV.

The collaboration of experimentalist and theorist communities has played
an important role in understanding current data, and it will be essential
also in future.

The LHC has shown that hadron colliders can do not only new physics searches,
but also precision physics, as already demonstrated by the experiments at
Tevatron.

As of today, we analysed less than 10\% of the dataset expected from LHC,
and at a centre-of-mass energy about half of the energy of future LHC runs,
starting from 2015. 
Furthermore, ultimate LHC precision studies to probe for new physics effects
at the TeV scale can be performed with the luminosity upgrade of the machine 
and of the detectors.

With the new start of LHC after the long shutdown of 2012-2014, we are
entering a new era in HEP, with the LHC acting as a ``portal" to a 
possible new world.


\end{document}

%% file: econfmacros.tex
\def\beq{\begin{equation}}
\def\eeq#1{\label{#1}\end{equation}}
\def\eeqn{\end{equation}}

\def\beqa{\begin{eqnarray}}
\def\eeqa#1{\label{#1}\end{eqnarray}}
\def\eeqan{\end{eqnarray}}

\def\sst{\scriptscriptstyle}

\let\bar=\overbar

\def\Dslash{\not{\hbox{\kern-4pt $D$}}}
\def\dslash{\not{\hbox{\kern-2pt $\del$}}}

\def\msb{{\bar{\ssstyle M \kern -1pt S}}}

